\documentclass[twocolumn,aps,prc,superscriptaddress,showpacs,floatfix]{revtex4}
\usepackage{amssymb}
\usepackage{amsmath}
\usepackage{graphicx}

\begin{document}

\title{Nuclear symmetry potential in the relativistic impulse approximation}
\author{Zeng-Hua Li}
\affiliation{Institute of Theoretical Physics, Shanghai Jiao Tong University, Shanghai
200240, China}
\author{Lie-Wen Chen}
\affiliation{Institute of Theoretical Physics, Shanghai Jiao Tong University, Shanghai
200240, China}
\affiliation{Center of Theoretical Nuclear Physics, National Laboratory of Heavy Ion
Accelerator, Lanzhou 730000, China}
\author{Che Ming Ko}
\affiliation{Cyclotron Institute and Physics Department, Texas A\&M University, College
Station, Texas 77843-3366, USA}
\author{Bao-An Li}
\affiliation{Department of Physics, Texas A\&M University-Commerce, Commerce, Texas
75429-3011, USA}
\author{Hong-Ru Ma}
\affiliation{Institute of Theoretical Physics, Shanghai Jiao Tong University, Shanghai
200240, China}
\date{\today }

\begin{abstract}
Using the relativistic impulse approximation with the Love-Franey \textsl{NN}
scattering amplitude developed by Murdock and Horowitz, we investigate the
low-energy ($100$ MeV$\leq E_{\mathrm{kin}}\leq 400$ MeV) behavior of the
nucleon Dirac optical potential, the Schr\"{o}dinger-equivalent potential,
and the nuclear symmetry potential in isospin asymmetric nuclear matter. We
find that the nuclear symmetry potential at fixed baryon density decreases
with increasing nucleon energy. In particular, the nuclear symmetry
potential at saturation density changes from positive to negative values at
nucleon kinetic energy of about $200$ MeV. Furthermore,the obtained energy
and density dependence of the nuclear symmetry potential is consistent with
those of the isospin- and momentum-dependent MDI interaction with $x=0$,
which has been found to describe reasonably both the isospin diffusion data
from heavy-ion collisions and the empirical neutron-skin thickness of $^{208}
$Pb.
\end{abstract}

\maketitle

\section{introduction}

The energy dependence of the nuclear symmetry potential, i.e., the isovector
part of nucleon mean-field potential in asymmetric nuclear matter, has
recently attracted much attention \cite%
{91bomb,97ulrych,das03,li04a,li04b,chen04,rizzo04,fuchs04,mazy04,chen05,li05xmed,05behera,baran05,samma05,zuo05,fuchs05,fuchs05prc,rizzo05,mazy06}%
. Its knowledge together with that of the density dependence of the nuclear
symmetry energy are important for understanding not only the structure of
radioactive nuclei and the reaction dynamics induced by rare isotopes but
also many critical issues in astrophysics \cite{ireview98,ibook}. Various
microscopic and phenomenological models, such as the relativistic
Dirac-Brueckner-Hartree-Fock (DBHF) \cite%
{97ulrych,fuchs04,mazy04,samma05,fuchs05,fuchs05prc,mazy06} and the
non-relativistic Brueckner-Hartree-Fock (BHF) \cite{91bomb,zuo05} approach,
the relativistic mean-field theory based on nucleon-meson interactions \cite%
{baran05}, and the non-relativistic mean-field theory based on Skyrme-like
interactions \cite{das03,05behera}, have been used to study the nuclear
symmetry potential, but the predicted results were found to vary widely.
While most models predict a decreasing nuclear symmetry potential with
increasing nucleon momentum, albeit at different rates, a few nuclear
effective interactions used in some of the models lead to an opposite
conclusion.

Using the relativistic Dirac optical potential obtained from the
relativistic impulse approximation (RIA) \cite%
{mcneil83prc,mcneil83,shepard83,arnold79,clark83,miller83,horowitz85,murdock87,tjon85,ott88,jin93,toki01}
with the empirical nucleon-nucleon (\textsl{NN}) amplitude calculated by
McNeil, Ray, and Wallace (MRW) \cite{mcneil83prc,mcneil83,shepard83}, which
works well for elastic nucleon-nucleus scattering at medium and high
energies (above $500$ MeV), three of present authors \cite{chen05ria} have
recently studied the high-energy behavior of the nuclear symmetry potential
in asymmetric nuclear matter. It was found that for nucleons at high
energies, the symmetry potential at fixed baryon density is essentially
constant and is slightly negative below nuclear density of about $\rho =0.22$
fm$^{-3}$ but increases almost linearly to positive values at high densities.

A nice feature of RIA is that it permits very little phenomenological
freedom in deriving the nucleon Dirac optical potential in nuclear matter.
The basic ingredients in this method are the free \textsl{NN} invariant
scattering amplitude and the nuclear scalar and vector densities in nuclear
matter. This is in contrast to the relativistic DBHF approach, where
different approximation schemes and methods have been introduced for
determining the Lorentz and isovector structure of the nucleon self-energy 
\cite{97ulrych,fuchs04,mazy04,samma05,fuchs05,fuchs05prc,mazy06}. However,
the original RIA of MRW failed to describe spin observables at laboratory
energies lower than $500$ MeV\cite{ray85}, and its predicted oscillations in
the analyzing power in proton-\textrm{Pb} scattering at large angles were
also in sharp disagreement with experimental data \cite{drake85}. These
shortcomings are largely due to the implicit dynamical assumptions about the
relativistic \textsl{NN} interaction in the form of the Lorentz covariance 
\cite{adams84} and the somewhat awkward behavior under the interchange of
two particles \cite{horowitz85} as well as the omitted medium modification
due to the Pauli blocking effect. To solve these theoretical limitations at
lower energies, Murdock and Horowitz (MH) \cite{horowitz85,murdock87}
extended the original RIA to take into account following three improvements:
i) an explicit exchange contribution was introduced by fitting to the
relativistic \textsl{NN} scattering amplitude; ii) a pseudovector coupling
rather than a pseudoscalar coupling was used for the pion; iii) medium
modification from the Pauli blocking was included. With these improvements,
the RIA with free \textsl{NN} scattering amplitude was then able to
reproduce successfully measured analyzing power and spin rotation function
for all considered closed shell nuclei in proton scattering near $200$ MeV.
Particularly, the medium modification due to the Pauli blocking effect was
found to be essential in describing the spin rotation function for $^{208}$%
Pb at proton energy of $290 $MeV \cite{murdock87}.

Extending our previous work by using the generalized RIA of MH and the
nuclear scalar and vector densities from the relativistic mean-field theory,
we study in the present paper the low-energy ($100$ MeV$\leq E_{\mathrm{kin}%
}\leq 400$ MeV) behavior of the nucleon Dirac optical potential, the Schr%
\"{o}dinger-equivalent potential, and the nuclear symmetry potential in
isospin asymmetric nuclear matter. We find that for low energy nucleons the
nuclear symmetry potential at fixed nuclear density decreases with
increasing nucleon energy. In particular, the nuclear symmetry potential at
saturation density changes from positive to negative values at nucleon
kinetic energy around $200$ MeV. The resulting energy and density dependence
of the nuclear symmetry potential is further found to be consistent with the
isospin- and momentum-dependent MDI interaction with $x=0$ \cite%
{das03,chen05}, which has been constrained by the isospin diffusion data in
heavy-ion collisions and the empirical neutron-skin thickness of $^{208}$Pb 
\cite{chen05,li05xmed,steiner05nskin,chen05nskin}. Our results thus provide
an important consistency check for the energy dependence of the nuclear
symmetry potential in asymmetric nuclear matter.

The paper is organized as follows. In Section \ref{optical}, we briefly
review the generalized relativistic impulse approximation for the nuclear
optical potential. Results on the relativistic nuclear optical potential,
the resulting Schr\"{o}dinger-equivalent potential, and the nuclear symmetry
potential in asymmetric nuclear matter are presented in Section \ref{results}%
. A summary is given in Section \ref{summary}.

\section{The theoretical method}

\label{optical}

\subsection{The relativistic impulse approximation to the Dirac optical
potential}

In the relativistic impulse approximation, the momentum-space optical
potential in a spin saturated nucleus is given by 
\begin{equation}
{\tilde{U}}(\mathbf{q})=\frac{-4\pi ip_{\text{lab}}}{M}[F_{S}(q){\tilde{\rho}%
}_{S}(\mathbf{q})+\gamma _{0}F_{V}(q)\tilde{\rho}_{V}(\mathbf{q})],
\label{trhom}
\end{equation}%
where $F_{S}$ and $F_{V}$ are, respectively, the scalar and the zeroth
component of the vector \textsl{NN} scattering amplitude; ${\tilde{\rho}}%
_{S}(\mathbf{q})$ and $\tilde{\rho}_{V}(\mathbf{q})$ are corresponding
momentum-space nuclear densities; $p_{\text{lab}}$ and $M$ are,
respectively, the laboratory momentum and mass of an incident nucleon; $%
\gamma _{0}$ is a Dirac gamma matrix; and $\mathbf{q}$ is the momentum
transfer. The optical potential in the coordinator space is given by the
Fourier transformation of ${\tilde{U}}(\mathbf{q})$. In infinite nuclear
matter, the nuclear coordinate-space Dirac optical potential takes the
simple form \cite{mcneil83,chen05ria}: 
\begin{equation}
U=\frac{-4\pi ip_{\text{lab}}}{M}[F_{S0}\rho _{S}+\gamma _{0}F_{V0}\rho
_{V}],  \label{trhor}
\end{equation}%
where $F_{S0}\equiv F_{S}(q=0)$ and $F_{V0}\equiv F_{V}(q=0)$ are the 
\textsl{NN} forward scattering amplitudes while $\rho _{S}$ and $\rho _{V}$
are, respectively, the spatial scalar and vector densities of an infinite
nuclear matter.

The Dirac optical potential in Eq. (\ref{trhor}) is valid for nucleons at
high energies. With decreasing nucleon energy, medium modification due to
the Pauli blocking effect becomes important. As described in detail in Ref. 
\cite{murdock87}, the Dirac optical potential including the Pauli blocking
effect can be written as 
\begin{equation}
U_{\text{opt}}=\left[ 1-a_{i}(E_{\mathrm{kin}})\left( \frac{\rho _{B}}{\rho
_{0}}\right) ^{2/3}\right] U,  \label{noisopb}
\end{equation}%
where $\rho _{B}$ is the nuclear baryon density and $\rho _{0}=0.1934$ fm$%
^{-3}$. The parameters $a_{i}(E_{\mathrm{kin}})$ denote the Pauli blocking
factors for each energy $E_{\mathrm{kin}}$ and are given in Table II of Ref. 
\cite{murdock87}. Although there are still many open questions on the role
of medium modification \cite{murdock87}, the $\rho _{B}^{2/3}$ density
dependence of the Pauli blocking factor is consistent with the phase-space
consideration for isotropic scattering \cite{chen01}. For nucleon scattering
in isospin asymmetric nuclear matter, the Pauli blocking effect becomes
different for protons and neutrons. Following Ref. \cite{chen01}, we
introduce an isospin-dependent Pauli blocking factor and obtain following
different Dirac optical potentials for protons and neutrons: 
\begin{eqnarray}
U_{\text{opt}}^{n(p)} &=&\left\{ 1-a_{i}(E_{\mathrm{kin}})\left[ \frac{%
(2\rho _{n(p)})^{2/3}+0.4(2\rho _{p(n)})^{2/3}}{1.4\rho _{0}^{2/3}}\right]
\right\}  \notag \\
&\times &U^{n(p)}.  \label{isopb}
\end{eqnarray}%
Obviously, Eq. (\ref{isopb}) reduces to Eq. (\ref{noisopb}) in symmetric
nuclear matter with $\rho _{n}=\rho _{p}$.

\subsection{Nuclear scalar densities}

To evaluate the Dirac optical potential for nucleons in RIA, we need to know
the nuclear scalar and vector densities. They can be determined from the
relativistic mean-field (RMF) model \cite{rmf}. Currently, there are many
different versions for the RMF model and they mainly include the non-linear
models \cite{rmf,reinhard89,ring96}, the models with density dependent
meson-nucleon couplings \cite{fuchs95,shen97,typel99,hofmann01}, and the
point-coupling models \cite%
{furnstahl97,reinhard02,reinhard04a,reinhard04b,weise04}. As in Ref. \cite%
{chen05ria}, we use in the present work the non-linear RMF model with a
Lagrangian density that includes the nucleon field $\psi $, the
isoscalar-scalar meson field $\sigma $, the isoscalar-vector meson field $%
\omega $, the isovector-vector meson field $\rho $, and the isovector-scalar
meson field $\delta $, i.e., 
\begin{eqnarray}
&&{\mathcal{L}}(\psi ,\sigma ,\bm{\omega},\bm{\rho},\delta )=\bar{\psi}\left[
\bm{\gamma}_{\mu }(i\partial ^{\mu }-g_{\omega }\bm{\omega}^{\mu
})-(M-g_{\sigma }\sigma )\right] \psi  \notag \\
&&+\frac{1}{2}(\partial _{\mu }\sigma \partial ^{\mu }\sigma -m_{\sigma
}^{2}\sigma ^{2})-\frac{1}{4}\bm{\omega}_{\mu \nu }\bm{\omega}^{\mu \nu }+%
\frac{1}{2}m_{\omega }^{2}\bm{\omega}_{\mu }\bm{\omega}^{\mu }  \notag \\
&&-\frac{1}{3}b_{\sigma }M{(g_{\sigma }\sigma )}^{3}-\frac{1}{4}c_{\sigma }{%
(g_{\sigma }\sigma )}^{4}+\frac{1}{4}c_{\omega }{(g_{\omega }^{2}\bm{\omega}%
_{\mu }\bm{\omega}^{\mu })}^{2}  \notag \\
&&+\frac{1}{2}(\partial _{\mu }\delta \partial ^{\mu }\delta -m_{\delta }^{2}%
{\delta }^{2})+\frac{1}{2}m_{\rho }^{2}\bm{\rho}_{\mu }.\bm{\rho}^{\mu }-%
\frac{1}{4}\bm{\rho}_{\mu \nu }.\bm{\rho}^{\mu \nu }  \notag \\
&&+\frac{1}{2}(g_{\rho }^{2}\bm{\rho}_{\mu }.\bm{\rho}^{\mu })(\Lambda
_{S}g_{\sigma }^{2}\sigma ^{2}+\Lambda _{V}g_{\omega }^{2}\bm{\omega}_{\mu }%
\bm{\omega}^{\mu })  \notag \\
&&-g_{\rho }\bm{\rho}_{\mu }\bar{\psi}\gamma ^{\mu }\bm{\tau}\psi +g_{\delta
}\delta \bar{\psi}\bm{\tau}\psi \;,  \label{lag}
\end{eqnarray}%
where the antisymmetric field tensors $\omega _{\mu \nu }$ and $\rho _{\mu
\nu }$ are given by $\bm{\omega}_{\mu \nu }\equiv \partial _{\nu }\bm{\omega}%
_{\mu }-\partial _{\mu }\bm{\omega}_{\nu }$ and $\text{ }\bm{\rho}_{\mu \nu
}\equiv \partial _{\nu }\bm{\rho}_{\mu }-\partial _{\mu }\bm{\rho}_{\nu }$,
respectively, and the symbols used in Eq. (\ref{lag}) have their usual
meanings. The above Lagrangian density is quite general in the non-linear
RMF model and allows us to use many presently popular parameter sets.

In Ref. \cite{chen05ria}, three typical parameter sets were used to evaluate
the scalar densities of neutrons and protons in asymmetric nuclear matter,
and they are the very successful NL3 model \cite{lala97}, the Z271v model,
and the HA model. The Z271v model has been used to study the neutron skin of
heavy nuclei and the properties of neutron stars \cite{horowitz01}, while
the HA model includes the isovector-scalar meson field $\delta $ and fits
successfully some results calculated with the more microscopic DBHF approach 
\cite{bunta03}. As shown in Ref. \cite{chen05ria}, the scalar densities of
neutrons and protons in asymmetric nuclear matter obtained from these three
parameter sets are similar at low baryon densities but become different for $%
\rho _{B}\gtrsim 0.25$ fm$^{-3}$, with Z271v giving a larger and NL3 a
smaller scalar density than that from the parameter set HA. For $\rho
_{B}\lesssim 0.25$ fm$^{-3}$, the proton and neutron scalar densities from
these three RMF models are also consistent with those from the RMF model
with density-dependent meson-nucleon couplings \cite{hofmann01}. The real
and imaginary parts of the scalar potential at higher densities ($\rho
_{B}\gtrsim 0.25$ fm$^{-3}$) thus depend on the interactions used in
evaluating the nuclear scalar density and have, therefore, large
uncertainties. In the present work, we only use the HA parameter set and
focus on nuclear densities smaller than $\rho _{B}\lesssim 0.25$ fm$^{-3}$
where the scalar densities of neutrons and protons in asymmetric nuclear
matter are essentially independent of the model parameters.

\subsection{Nuclear symmetry potential}

In the Dirac spinor space of the projectile nucleon, the optical potential $%
U_{\text{opt}}$ is a $4\times 4$ matrix and can be expressed in terms of a
scalar $U_{S}^{\mathrm{tot}}$ and a vector $U_{0}^{\mathrm{tot}}$ piece: 
\begin{equation}
U_{\text{opt}}=U_{S}^{\mathrm{tot}}+\gamma _{0}U_{0}^{\mathrm{tot}}.
\end{equation}
Expressing $U_{S}^{\mathrm{tot}}$ and $U_{0}^{\mathrm{tot}}$ in terms of
their real and imaginary parts, i.e., 
\begin{equation}
U_{S}^{\mathrm{tot}}=U_{S}+iW_{S},\text{ \ }U_{0}^{\mathrm{tot}%
}=U_{0}+iW_{0},
\end{equation}
a \textquotedblleft Schr\"{o}dinger-equivalent potential \textquotedblright\
(SEP) can be obtained from the Dirac optical potential \cite%
{brock78,jaminon80}: 
\begin{equation}
U_{\text{SEP}}=U_{S}^{\mathrm{tot}}+U_{0}^{\mathrm{tot}}+\frac{1}{2M}(U_{S}^{%
\mathrm{tot2}}-U_{0}^{\mathrm{tot2}})+\frac{U_{0}^{\mathrm{tot}}}{M} E_{%
\mathrm{kin}},  \label{SEP}
\end{equation}%
We note that solving the Schr\"{o}dinger equation with the SEP gives the
same bound-state energy eigenvalues and elastic phase shifts as the solution
of the upper component of the Dirac spinor in the Dirac equation using
corresponding Dirac optical potential.

The real part of SEP is then given by 
\begin{equation}
\text{Re}(U_{\text{SEP}})=U_{S}+U_{0}+\frac{1}{2M}%
[U_{S}^{2}-W_{S}^{2}-(U_{0}^{2}-W_{0}^{2})]+\frac{U_{0}}{M}E_{\mathrm{kin}}.
\label{ReSEP}
\end{equation}%
Above equation corresponds to the nuclear mean-field potential in
non-relativistic models \cite{jaminon89,fuchs05} and allows us to obtain the
following nuclear symmetry potential, i.e., the so-called Lane potential 
\cite{lane62}: 
\begin{equation}
U_{\text{sym}}=\frac{\text{Re}(U_{\text{SEP}})_{n}-\text{Re}(U_{\text{SEP}%
})_{p}}{2\alpha },
\end{equation}%
where Re$(U_{\text{SEP}})_{n}$ and Re$(U_{\text{SEP}})_{p}$ are,
respectively, the real part of SEP for neutrons and protons. The isospin
asymmetry $\alpha $ is defined as $\alpha =\frac{\rho _{n}-\rho _{p}}{\rho
_{n}+\rho _{p}}$with $\rho _{n}$ and $\rho _{p}$ denoting the neutron and
proton densities, respectively.

\section{results}

\label{results}

\subsection{Relativistic \textsl{NN} scattering amplitude}

\begin{figure}[th]
\includegraphics[scale=0.78]{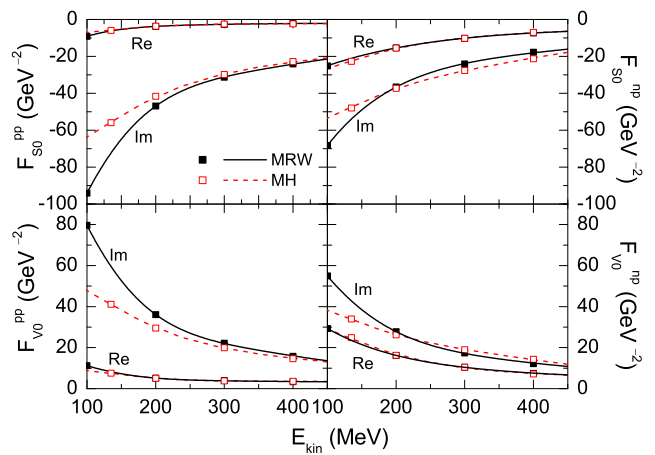}
\caption{{\protect\small (Color online) The scalar and vector parts of the
free \textsl{NN} forward scattering amplitudes }$F_{S0}^{pp}${\protect\small %
, }$F_{S0}^{np}${\protect\small , }$F_{V0}^{pp}${\protect\small , }and $%
F_{V0}^{np}${\protect\small \ at nucleon kinetic energies }$E_{\mathrm{kin}%
}=135${\protect\small \ MeV, }$200${\protect\small \ MeV, }$300$%
{\protect\small \ MeV and }$400${\protect\small \ MeV (open squares) from
the RIA of MH. Dashed lines are polynomial fits to the energy dependence of
the \textsl{NN} scattering amplitude. Corresponding results from the
original RIA of MRW are shown by solid squares and lines.}}
\label{FFSV}
\end{figure}

Based on the generalized RIA of MH with the Love-Frany \textsl{NN}
scattering amplitudes \cite{love81}, we have evaluated the scalar and vector
parts of the \textsl{NN} forward scattering amplitudes $F_{S0}^{pp}$, $%
F_{S0}^{np}$, $F_{V0}^{pp}$, and $F_{V0}^{np}$ at nucleon kinetic energies $%
E_{\mathrm{kin}}=135$ MeV, $200$ MeV, $300$ MeV and $400$ MeV for which the
parameters can be found explicitly in Refs. \cite{horowitz85,murdock87}, and
the resulting values are shown by open squares in Fig. \ref{FFSV}. To obtain
continuous and smooth results for the \textsl{NN} scattering amplitude and
other quantities in the following, we have made polynomial fits to the
energy dependence of the \textsl{NN} scattering amplitude, and the results
are shown by dashed lines in Fig. \ref{FFSV}. For comparison, we also
include corresponding results from the original RIA of MRW by solid squares
and lines. It is seen that for both proton-proton and proton-neutron
scattering, the real parts of corresponding amplitudes in the two approaches
are in good agreement with each other. However, for the imaginary parts of
the amplitudes, the strength of the scalar and vector amplitudes from the
RIA of MH displays much weaker energy dependence for both proton-proton and
proton-neutron scattering at the energies $E_{\mathrm{kin}}\leq 300$ MeV.
Since the imaginary part of the amplitude just corresponds to the real part
of the Dirac optical potential as shown in Eq. (\ref{trhor}), above
differences between the original RIA of MRW and the generalized RIA of MH
thus lead to different behavior of the Dirac optical potential at low
energies.

\subsection{Relativistic Dirac optical potential}

\begin{figure}[th]
\includegraphics[scale=0.8]{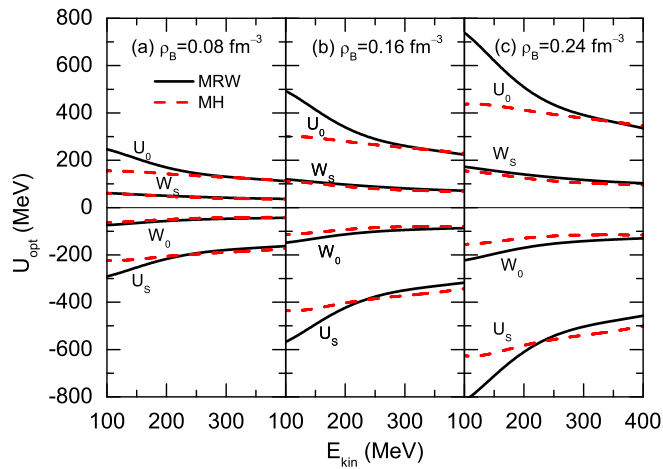}
\caption{{\protect\small (Color online) Energy dependence of the real and
imaginary parts of the scalar and vector optical potentials in symmetric
nuclear matter for different baryon densities }$\protect\rho _{B}$%
{\protect\small , with MH and MRW scattering amplitudes.}}
\label{OPden}
\end{figure}

With free \textsl{NN} forward scattering amplitudes of MH and MRW as well as
the neutron and proton scalar and vector densities obtained from the RMF
theory with the parameter set HA, we have investigated the real and
imaginary parts of the scalar and vector Dirac optical potentials for
nucleons in symmetric nuclear matter as functions of nucleon energy. In Fig. %
\ref{OPden}, the energy dependence of the Dirac optical potential is
depicted at three nucleon densities $\rho _{\text{B}}=0.08$ fm$^{-3}$ (panel
(a)), $0.16$ fm$^{-3}$ (panel (b)), and $0.24$ fm$^{-3}$ (panel (c)). In
each panel, we give the scalar and vector optical potentials based on the
generalized amplitudes of MH and the original amplitudes of MRW. In
calculating the Dirac optical potential from the RIA of MH, we have included
the Pauli blocking effect as well as the modifications from using the
pseudovector coupling for pion and the exchange term contribution. For all
densities we are considering, the energy dependence of the scalar and vector
optical potentials from the RIA of MH are significantly reduced compared
with those from the original RIA of MRW, especially for the real part at low
energies. Furthermore, the difference between the two becomes larger with
increasing density. These results thus demonstrate clearly the importance of
the medium modifications introduced in the RIA of MH for nucleons at low
energies. We note that for all three considered densities, the RIA of MH
generates, on the other hand, a similar systematic difference or isospin
splitting in the Dirac optical potentials for protons and neutrons in
asymmetric nuclear matter as in the original RIA of MRW \cite{chen05ria}. In
particular, the neutron exhibits a stronger real but weaker imaginary scalar
and vector potentials in neutron-rich nuclear matter.

\subsection{Schr\"{o}dinger-equivalent optical potential}

\begin{figure}[th]
\includegraphics[scale=0.85]{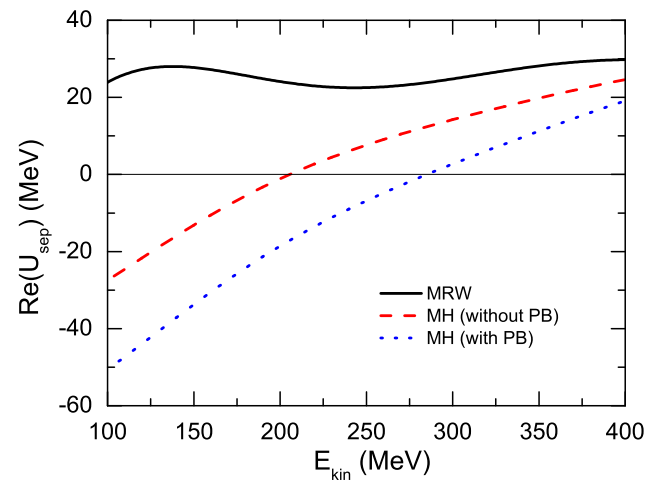}
\caption{{\protect\small (Color online) Energy dependence of the real part
for the nucleon Schr\"{o}dinger-equivalent potential at normal density in
symmetric nuclear matter, from the original RIA of MRW and from the RIA of
MH with and without Pauli blocking correction.}}
\label{Usep}
\end{figure}

Before discussing the nuclear symmetry potential, we first show in Fig. \ref%
{Usep} the real part of the nucleon Schr\"{o}dinger-equivalent potential in
symmetric nuclear matter at normal density obtained from above Dirac optical
potential. Because of uncertainties in the medium modification due to the
Pauli blocking effect at low energies, results both with (dotted line) and
without Pauli blocking (dashed line) corrections based on the MH free 
\textsl{NN} scattering amplitudes are shown. For comparison, we also show
the real part of the nucleon Schr\"{o}dinger-equivalent potential from the
original RIA of MRW (solid line). The nucleon Schr\"{o}dinger-equivalent
potential from the original RIA of MRW is seen to be always positive at
considered energy range of $E_{\mathrm{kin}}=100\sim 400$ MeV. Including the
pseudovector coupling and exchange term corrections in the RIA of MH (dashed
line), the behavior of the resulting Schr\"{o}dinger-equivalent potential as
a function of energy is significantly improved, varying from $-27$\ MeV at $%
E_{\mathrm{kin}}=100$ MeV to $0$\ MeV at $E_{\mathrm{kin}}\thickapprox 200$
MeV and then continues to increase monotonously as the nucleon energy
increases. This improvement is due to the fact that the pseudovector
coupling and exchange term corrections lead to a smaller strength of the
imaginary scalar and vector \textsl{NN} forward scattering amplitudes while
keep their sum roughly unchanged as shown in Fig. \ref{FFSV}. From Eq. (\ref%
{trhor}), therefore, the term $U_{S}+U_{0}$ does not change while the last
two terms in Eq. (\ref{ReSEP}) are reduced strongly and thus a smaller Schr%
\"{o}dinger-equivalent potential is obtained. When the Pauli blocking effect
is further taken into account, the resulting Schr\"{o}dinger-equivalent
potential is seen to be more attractive at the whole energy range considered
here. At high enough energy, the Schr\"{o}dinger-equivalent potentials from
above three approaches become similar as expected since effects from both
Pauli blocking and exchange contribution play minor role at high energies.

We note that with momentum/energy independent scalar and vector potentials
from the RMF calculation, the nucleon Schr\"{o}dinger-equivalent potential
in symmetric nuclear matter at normal nuclear density exhibits already a
linear energy dependence according to Eq. (\ref{ReSEP}), with a change from
negative to positive values typically at kinetic energies between about $200$
MeV and $500$ MeV depending on the model parameters \cite{weber92}.
Experimental data from the global relativistic optical-model analysis also
indicate that the nucleon Schr\"{o}dinger-equivalent potential in symmetric
nuclear matter at normal nuclear density changes from negative to positive
values around $200$ MeV, although with large uncertainties, as mentioned in
Ref. \cite{weber92}.

\subsection{Nuclear symmetry potential}

\begin{figure}[th]
\includegraphics[scale=1.1]{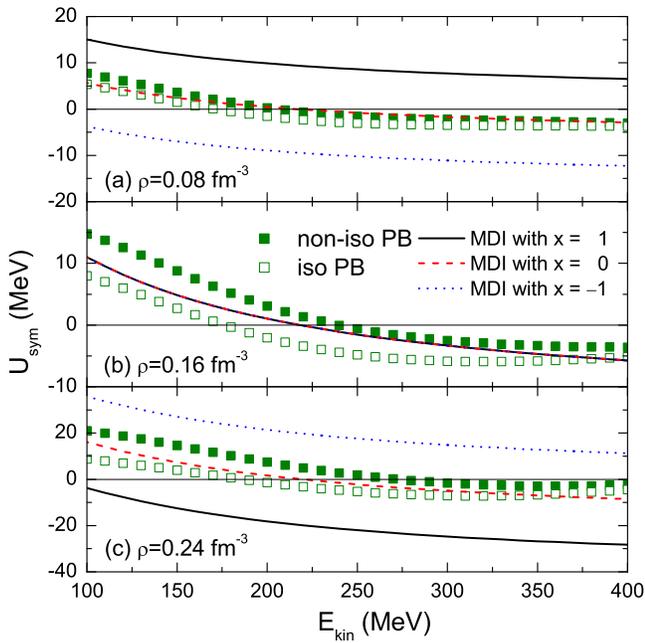}
\caption{{\protect\small (Color online) Energy dependence of the nuclear
symmetry potential from the RIA of MH with isospin-dependent (open squares)
and isospin-independent (solid squares) Pauli blocking corrections, as well
as the phenomenological MDI interaction with }$x=1${\protect\small \ (solid
line), }$0${\protect\small \ (dashed line)}$,${\protect\small \ and }$-1$%
{\protect\small \ (dotted line) at fixed baryon densities of }$\protect\rho %
_{B}=0.08${\protect\small \ \ fm}$^{-3}${\protect\small \ (a), }$0.16$%
{\protect\small \ fm}$^{-3}${\protect\small \ (b), and }$0.24$%
{\protect\small \ fm}$^{-3}${\protect\small \ (c).}}
\label{UsymEkin}
\end{figure}

For the nuclear symmetry potential based on the scattering amplitudes of MH,
we show in Fig. \ref{UsymEkin} its energy dependence for both cases of using
isospin-dependent (Eq. (\ref{isopb})) and isospin-independent Pauli blocking
(Eq. (\ref{noisopb})) corrections at fixed baryon densities of $\rho
_{B}=0.08$ fm$^{-3}$ (panel (a)), $0.16$ fm$^{-3}$ (panel (b)), and $0.24$ fm%
$^{-3}$ (panel (c)). Also shown are results from the phenomenological
parametrization of the isospin- and momentum-dependent nuclear mean-field
potential, i.e., the MDI interaction with $x=-1$, $0$, and $1$, which has
recently been extensively used in the transport model for studying isospin
effects in intermediate-energy heavy ion collisions induced by neutron-rich
nuclei \cite{li04b,chen04,chen05,li05xmed,li05pion,li06,yong061,yong062}. In
the MDI interaction, the single nucleon potential in asymmetric nuclear
matter with isospin asymmetry $\alpha $ is expressed by \cite%
{das03,li04b,chen04,chen05,li05xmed} 
\begin{eqnarray}
&&U(\rho ,\alpha ,\mathbf{{p},\tau ,{r})}  \notag \\
&=&\left( -95.98-x\frac{2B}{\sigma +1}\right) \frac{\rho _{\tau ^{\prime }}}{%
\rho _{0}}  \notag \\
&&+\left( -120.57+x\frac{2B}{\sigma +1}\right) \frac{\rho _{\tau }}{\rho _{0}%
}  \notag \\
&&+B\left( \frac{\rho }{\rho _{0}}\right) ^{\sigma }(1-x\alpha ^{2})-8\tau x%
\frac{B}{\sigma +1}\frac{\rho ^{\sigma -1}}{\rho _{0}^{\sigma }}\alpha \rho
_{\tau ^{\prime }}  \notag \\
&&+\frac{2C_{\tau ,\tau }}{\rho _{0}}\int d^{3}\mathbf{p}^{\prime }\frac{%
f_{\tau }(\mathbf{{r},{p}^{\prime })}}{1+(\mathbf{p}-\mathbf{p}^{\prime
})^{2}/\Lambda ^{2}}  \notag \\
&&+\frac{2C_{\tau ,\tau ^{\prime }}}{\rho _{0}}\int d^{3}\mathbf{p}^{\prime }%
\frac{f_{\tau ^{\prime }}(\mathbf{{r},{p}^{\prime })}}{1+(\mathbf{p}-\mathbf{%
p}^{\prime })^{2}/\Lambda ^{2}}.  \label{mdi}
\end{eqnarray}%
In the above $\tau =1/2$ ($-1/2$) for neutrons (protons) and $\tau \neq \tau
^{\prime }$; $\sigma =4/3$; and $f_{\tau }(\mathbf{{r},{p})}$ is the
phase-space distribution function at coordinate $\mathbf{r}$ and momentum $%
\mathbf{p}$. The parameters $B$, $C_{\tau ,\tau }$, $C_{\tau ,\tau ^{\prime
}}$ and $\Lambda $ are determined by fitting the momentum-dependence of $%
U(\rho ,\alpha ,\mathbf{p},\tau ,\mathbf{r})$ to that predicted by the Gogny
Hartree-Fock and/or the Brueckner-Hartree-Fock (BHF) calculations \cite%
{91bomb}, the saturation properties of symmetric nuclear matter and the
symmetry energy of $31.6$ MeV at normal nuclear matter density $\rho
_{0}=0.16$ fm$^{-3}$ \cite{das03}. The incompressibility $K_{0}$ of
symmetric nuclear matter at $\rho _{0}$ is set to be $211$ MeV. The
different $x$ values in the MDI interaction are introduced to vary the
density dependence of the nuclear symmetry energy while keeping other
properties of the nuclear equation of state fixed \cite{chen05}. We note
that the energy dependence of the symmetry potential from the MDI
interaction is consistent with the empirical Lane potential at normal
nuclear matter density and low nucleon energies \cite{li04a}.

It is seen from Fig. \ref{UsymEkin} that at fixed baryon density, the
nuclear symmetry potential generally decreases with increasing nucleon
energy. At low nuclear density ($\rho _{\text{B}}=0.08$ fm$^{-3}$), the
symmetry potentials from the RIA of MH with isospin-dependent and
isospin-independent Pauli blocking corrections are almost the same,
especially at energies higher than $E_{\mathrm{kin}}\geqslant 300$ MeV,
where the Pauli blocking correction is expected to be unimportant. The
isospin dependence of the Pauli blocking effect becomes, however, stronger
as nuclear density increases, and an appreciable difference in the resulting
symmetry potentials is seen. The difference disappears, however, for high
energy nucleons when the Pauli Blocking effect becomes negligible. It is
interesting to note that at normal density ($\rho _{\text{B}}=0.16$ fm$^{-3}$%
), the nuclear symmetry potential changes from positive to negative values
at nucleon kinetic energy around $200$ MeV, with the one using the
isospin-dependent Pauli blocking correction at a somewhat lower energy than
that using the isospin-independent Pauli blocking correction. Comparing with
results from the MDI interaction, the one with $x=0$ is seen in surprisingly
good agreement with the results of RIA by MH in the region of nuclear
densities and energies considered here. Although the MDI interaction with
different $x$ values give by construction same symmetry potential at normal
nuclear matter density as shown in Fig. \ref{UsymEkin}(b), the one with $x=0$
has been found to give reasonable descriptions of the data on the isospin
diffusion in intermediate energy heavy ion collisions and the neutron skin
thickness of $^{208}$Pb \cite{chen05,li05xmed,steiner05nskin,chen05nskin}.

\begin{figure}[th]
\includegraphics[scale=0.85]{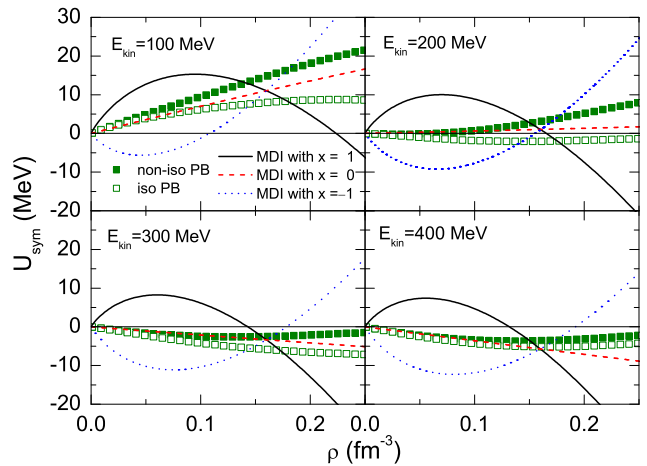}
\caption{{\protect\small (Color online) Density dependence of the nuclear
symmetry potential using RIA with isospin dependent and independent Pauli
blocking, as well as the results from the phenomenological interaction MDI
with} $x=-1,0,${\protect\small \ and }$1$ {\protect\small at nucleon kinetic
energies of }$E_{\mathrm{kin}}=100$ {\protect\small MeV, }$200$ 
{\protect\small MeV, }$300$ {\protect\small MeV and }$400$ {\protect\small %
MeV.}}
\label{UsymDen}
\end{figure}

The density dependence of the nuclear symmetry potential with
isospin-dependent and isospin-independent Pauli blocking corrections at
nucleon kinetic energies of $100$, $200$, $300$, $400$ MeV are shown in Fig. %
\ref{UsymDen} together with corresponding results from the MDI interaction
with $x=-1$, $0$, and $1$. It is clearly seen that the nuclear symmetry
potentials are always positive at lower nucleon kinetic energy of $E_{%
\mathrm{kin}}=100$ MeV while it may become positive or negative at $E_{%
\mathrm{kin}}=200$ MeV depending on if the Pauli blocking effect is isospin
dependent or not. At higher energies ($E_{\mathrm{kin}}=300$ and $400 $
MeV), the nuclear symmetry potential is always negative in the density
region considered here. These features are consistent with the results shown
in Fig. \ref{UsymEkin}. Compared with results from the MDI interaction, it
is seen that the nuclear symmetry potential from the generalized RIA of MH
reproduces nicely the results obtained from the MDI interaction with $x=0$
when $\rho _{\text{B}}\lesssim 0.2$ fm$^{-3}$ even for nucleon kinetic
energy as high as $400$ MeV. Moreover, in the energy region of $E_{\mathrm{%
kin}}=100\sim 300$ MeV, the nuclear symmetry potential from MDI interaction
with $x=0$ always lies between the results from the RIA of MH with
isospin-dependent and isospin-independent Pauli blocking corrections. On the
other hand, the MDI interaction with both $x=-1$ and $1$\ display very
different density dependence from the results using the RIA of MH.

\section{summary}

\label{summary}

Based on the generalized relativistic impulse approximation of MH and the
scalar and vector densities from the relativistic mean-field theory with the
parameter set HA, we have studied the low-energy behavior of the nuclear
symmetry potential in asymmetric nuclear matter. In the relativistic impulse
approximation of MH, the low energy behavior of the Dirac optical potential
has been significantly improved by including the pseudovector coupling for
pion, the exchange contribution, and medium modification due to the Pauli
blocking effect. We find that compared with results from the original RIA of
MRW, the generalized RIA of MH gives essentially identical real parts of the
scalar and vector amplitudes for both proton-proton and neutron-proton
scattering but significantly reduced strength in their imaginary parts at
low energies $E_{\mathrm{kin}}\leq 300$ MeV. These improvements in the RIA
of MH modify the real scalar and vector Dirac optical potentials at lower
energies and make the resulting energy dependence of the Schr\"{o}%
dinger-equivalent potential and nuclear symmetry potential more reasonable.

At saturation density, the nuclear symmetry potential is found to change
from positive to negative values at nucleon kinetic energy of about $200$
MeV. This is a very interesting result as it implies that the proton
(neutron) feels an attractive (repulsive) symmetry potential at lower
energies but repulsive (attractive) symmetry potential at higher energies in
asymmetric nuclear matter. Adding also the repulsive Coulomb potential, a
high energy proton in asymmetric nuclear matter thus feels a very stronger
repulsive potential. This behavior of the nuclear symmetry potential can be
studied in intermediate and high energy heavy-ion collisions that are
induced by radioactive nuclei, e.g., by measuring two-nucleon correlation
functions \cite{chen-nn} and light cluster production \cite{chen-cluster} in
these collisions.

Comparing the energy and density dependence of the nuclear symmetry
potential from the RIA of MH with that from the MDI interaction indicates
that results from the MDI interaction with $x=0$ are in good agreement with
those from the RIA of MH. For baryon density less than $0.25$ fm$^{-3}$ and
nucleon energy less than $400$ MeV as considered in the present work, the
nuclear symmetry potential from the MDI interaction with $x=0$ lies
approximately between the two results from the RIA of MH with
isospin-dependent and isospin-independent Pauli blocking corrections. This
provides a strong evidence for the validity of the MDI interaction with $x=0$
in describing both the isospin diffusion data in intermediate energy heavy
ion collisions and the neutron skin thickness data for $^{208}$Pb.

The results presented in present work thus provide an important consistency
check for the energy/momentum dependence of the nuclear symmetry potential
in asymmetric nuclear matter, particularly the momentum dependent MDI
interaction with $x=0$, which is an essential input to the isospin-dependent
transport model \cite{li04b,baran05,li05xmed} in studying heavy-ion
collisions induced by radioactive nuclei at intermediate energies. They are
also useful in future studies that extend the Lorentz-covariant transport
model \cite{ko87,mosel92} to include explicitly the isospin degrees of
freedom.

\begin{acknowledgments}
This work was supported in part by the Post-doctors Research Award Fund of
China, the National Natural Science Foundation of China under Grant Nos.
10334020, 10575071 and 10675082, MOE of China under project NCET-05-0392,
Shanghai Rising-Star Program under Grant No. 06QA14024, the US National
Science Foundation under Grant Nos. PHY-0457265, PHY-0354572 and
PHY-0456890, the Welch Foundation under Grant No. A-1358, and the
NASA-Arkansas Space Grants Consortium Award ASU15154.
\end{acknowledgments}

\end{document}